\documentstyle[12pt, leqno]{article}
\def\appendixa{
\vskip 1cm
{\Large\bf Appendix A}
\vskip 1cm
\par
\setcounter{equation}{0}
\def\theequation{A.\arabic{equation}}
}
\def\appendixb{
\vskip 1cm
{\Large\bf Appendix B}
\vskip 1cm
\par
\setcounter{equation}{0}
\def\theequation{B.\arabic{equation}}
}

\begin{document}
\renewcommand{\theequation}{\thesection.\arabic{equation}}
\newcommand{\rl}{{\cal L}}
\thispagestyle{empty}
{\hfill \bf DFT-US-3/94}
\vspace*{0.5cm}
\begin{center}
{\Large \bf Dynamics of relativistic particle  \\ with  Lagrangian
 dependent on acceleration}
\vskip0.6cm
{\large  V.\ V.\ Nesterenko}\footnote{{\it E-mail address:}
     NESTR@theor.jinrc.dubna.su}
 \vskip 0.3cm

{ \it   Bogolubov Laboratory of Theoretical Physics,
 \\ Joint Institute for Nuclear Research \\
     Dubna, SU 141980 Russia}  \\
\vskip 0.3cm
and\\
\vskip 0.3cm
{\large A. \ Feoli, \  \  \ G.\ Scarpetta}\footnote{{\it E-mail address:}
SCARPETTA@salerno.infn.it}
              \\
\vskip 0.3cm
{ \it Dipartimento  di  Fisica Teorica e   s.m.s.a. --
Universit\`a di Salerno
 \\
     84081 Baronissi (SA), Italia \\
     Istituto Nazionale di Fisica Nucleare -- Sezione di Napoli \\
    International Institute for Advanced Scientific Studies
 -- Vietri sul Mare (SA)}\\
[1.4cm]

                         {\large \bf  Abstract }   \\[0.4cm]
   \end{center}

     Models of      relativistic   particle  with
Lagrangian ${\cal L}(k_1)$, depending on the   curvature
 of  the  worldline  $k_1$,
are  considered.  By making use of the Frenet basis,
the equations of motion  are  reformulated  in  terms  of  the
principal curvatures  of the worldline.  It
is shown that for arbitrary Lagrangian function ${\cal L}(k_1)$
these equations are
completely integrable,   i.e.,  the  principal  curvatures  are
defined by integrals. The constants of integration are
the particle mass and its spin.  The  developed  method
is   applied to the study of a model  of
relativistic  particle  with  maximal proper  acceleration, whose
Lagrangian  is uniquely  determined    by a  modified form of the invariant
relativistic interval.  This model gives us an example of a consistent
relativistic  dynamics  obeying  the  principle of a superiorly limited
value of the acceleration, advanced recently.

     \newpage
     \setcounter{footnote}0
\section{Introduction}
 Lagrangians  depending on  higher derivatives of the curve coordinates have
been  recently considered in many problems:
 the null-dimensional (particle-like) version of the rigid string
[1 -- 7], the model of  boson-fermion transmutation in external Chern-Simons
field [8 -- 11], the polymer theory [12].

Lagrangians of this kind also
occur  in connection with a conjecture about  the existence  of  the
 maximal proper acceleration of a massive particle in arbitrary motion
 introduced by  E.R. Caianiello and coll. [13 -- 19]; they
    proposed  a   new geometric
approach to quantum mechanics in which the commutators
between coordinates and momenta are interpreted
 as the components of the curvature tensor
of the eight dimensional space--time tangent bundle TM, that,
in this scheme, acquires a metric structure.
 This fundamental constant of nature has been
also introduced starting from
the Heisenberg's uncertainty relations [20 -- 22 ],
 putting in evidence, in such a way,
the deeply  interconnection between  the maximal acceleration
and the extended nature of  particles, whose finite extension
 cannot be neglected without
introducing in quantum field theory troubles and divergencies connected with
the point-like approximation.
Recently these variational problems
also became interesting for mathematicians [23]. The list of references
is certainly incomplete; however, it illustrates the continuing interest
in the subject.

Investigation of these models, in the framework of the classical
variational calculus,    gives rise  to   very   complicated   nonlinear
differential equations of order $2p$
 ($p$  is the highest order of the derivatives in the
Lagrangian), which, practically,  cannot  be subject to
analysis.

     However, a   considerable  advance  can  be  achieved    by
applying the following basic result from the classical  differential
geometry [24, 25]: any smooth curve $x^\mu (s), \; \mu \,=\,0,1, \ldots ,
D-1$ in $D$-dimensional flat space-time is completely determined  (up
to   rotations   as  a  whole)  by  specifying  the $D-1$  principal
curvatures  $k_i(s), \; i\,=\,1,2, \ldots , D-1$, where
$s$ is  the  worldline  length.  Therefore,  one  can  try  to derive the
Euler-Lagrange equations  in  terms  of  the  principal   curvatures
$k_i(s),\; i\,=\,1,2,  \ldots  ,D-1$  of  the worldline,  rather
than in terms of the coordinates $x^\mu  (s)$.  The  order  of  the
differential equations for $k_i(s)$ will be necessarily lower than the
order  for $x^\mu (s)$, because the curvature $k_i(s)$ is a function
 of the derivatives of $x$ up to order $i\,+\,1$
 ($i\,=\,1,2, \ldots, D-1$).

     For  Lagrangians of the form $\rl (k_1,k_2,  \ldots , k_{D-1})$
the Euler-Lagrange  equations can be always reformulated
in terms of $k_i(s)$.  In the present paper it will be shown that the
equations of  motion  for $k_i(s)$, generated by arbitrary Lagrangian
function $\rl  (k_1(s))$,  are  always  integrable  by   quadratures.
Furthermore, the  constants of integration are expressed in terms of
the Noether invariants for the given Lagrangian.  If one assumes  that
the Lagrangian  $\rl  (k_1)$  defines  a  model  of    relativistic
particle, then the integration constants turn out to be the mass  and
spin of the particle.  In this case $k_1(s)$ is the proper acceleration
of the particle.

       In  Section  2, given an arbitrary Lagrangian $\rl (k_1)$,
we derive  the
equations of  motion in
terms of  the  principal  curvatures,  $k_i(s)$,  of the trajectory,
   by  making  use  of  the  Hamilton
principle.  The derivation  proposed by us  is  very  simple
and clear;      we shall only
apply the  standard Frenet equations describing the moving $D$-hedron
of the curve; the Griffiths' approach [23, 26], using the
Cartan formalism of exterior differential forms, is
similar but more involved.

In Section 3 the  equations  of  motion  for  principal
curvatures $k_i(s)$  are  integrated.  The constants of integration  are
expressed in  terms of the particle mass and its spin.

In Section 4
the efficiency of our formalism is illustrated by  investigating  the
model of    a   relativistic   particle   with  maximal proper
acceleration.

In  Section 5 we draw some conclusion and discuss shortly
the  results.  In  Appendix A  Lagrangians linear in
$k_1$ are considered; in Appendix B the Hamiltonian formalism for the
model of a relativistic particle  with  maximal proper  acceleration  is
presented.

\section{Euler-Lagrange equations in  terms  of  the  principal
curvatures}
     Let us consider a reparametrization-invariant action
     \begin{equation}
     S\,=\int \!\!{\cal L}(k_1)\,ds
     \end{equation}
with a  Lagrangian  ${\cal L}$ depending only on  the
first curvature  $k_1$  (on  the particle acceleration) of the worldline
 $x^\mu (s),  \;  \mu  \,=\,0,1,  \ldots  ,  D-1$  in
$D$-dimensional Minkowski  space.  Here  $ds  ^2 \,=\,dx^\mu dx_\mu$,
and $s$ is the natural parameter along the worldline  $x^\mu  (s)$
(its length)
     \begin{equation}
     \frac{dx^\mu}{ds}\,\frac{d x_\mu}{ds}\,=\,1\,{.}
     \end{equation}
   The Lorentz metric $\eta_{\mu \nu}$ with a signature $(+,-, \ldots ,-)$
   will be used. For shortening, the differentiation with
   respect to the natural parameter $s$ will be denoted by a dot.
 The first curvature (or simply curvature) is
   defined by
     \begin{equation}
     k_1^2(s) \,=\, -\,\ddot x_\mu\,\ddot x^\mu \,=\,- \ddot x^2\,{.}
     \end{equation}
     Following the Hamiltonian principle, we require
     $$
\delta S \,=\,\delta S_1 \,+\,\delta S_2 \,=
     $$
     \begin{equation}
    = \int \!\!  ds \,{\cal L}'(k_1)\,\delta k_1(s)\,+\,\int \!\!{\cal
L}(k_1)\,\delta ds \,=\,0\,{,}
     \end{equation}
      where the  variations  $\delta  k_1(s)$  and  $\delta  ds$ are
generated by the variation of the  worldline    $\delta  x^\mu
(s)$. The  prime  on the Lagrangian function ${\cal L}(k_1)$ denotes
the differentiation with respect to its argument $k_1$.

 At every point on the worldline $x^\mu (s)$ we can associate a Frenet
     basis [24, 25],  an  orthonormal $D$-hedron $ e^\mu _a(s)$,
(the Latin index is a $D$-hedron index,
 the Greek index specifies the space-time component),
which is formed by the
unit
time-like tangent vector
     \begin{equation}
   e^\mu _0(s) \,=\,\frac{dx^\mu}{ds}, \quad e_0^2\,=\,\dot x^2\,=\,1
     \end{equation}
  and by a set of $D-1$ unit space-like  vectors $e^\mu_j(s)$:
\begin{equation}
e^\mu_i\,e_{j\mu}\,=\,-\,\delta_{ij}, \quad e^\mu_0\,e_{j\mu}\,=\,0\,{,}
\quad 1\leq i,j \leq D-1\,{.}
\end{equation}
 Raising and lowering of indices are
made by the corresponding metric tensors
\begin{equation}
\eta_{a b}\,=\,\mbox{diag }(1,\,-1,\,\ldots ,\,-1),\quad
\eta_{\mu \nu}\,=\,\mbox{diag }(1,\,-1,\,\ldots ,\,-1)\,{.}
\end{equation}
The  orthonormality condition for Frenet basis is:
\begin{equation}
e^\mu _a \,e_{b \mu }\,=\,\eta_{a b}, \quad 0\,
\leq a, \,
b \,\leq D-1, \quad \mu \,=\,0,1, \ldots , D-1\,{.}
\end{equation}
We shall need the Frenet equations describing the change of the Frenet
basis under motion of its origin along the worldline
\begin{equation}
\dot e^\mu _a \,=\,\omega _a {}^b {} \,e^\mu _b, \quad
\omega _{a b}\,+\,\omega _{b a}\,=\,0\,{.}
\end{equation}
Nonzero elements of the matrix $\omega $ are determined by the principal
curvatures
\begin{equation}
\omega _{a ,\, a + 1}\,=\, -\, \omega _{a +1,\,
a}\,=\,k_{a +1}(s), \quad a \,=\,0,1,  \ldots ,D-2\,{.}
\end{equation}
Let us express the variation $\delta x^\mu (s)$ in terms of the Frenet
basis
$$
\delta x^\mu (s) \,=\, \varepsilon ^a (s)\,e^\mu _a (s),
$$
\begin{equation}
 \mu \,=\,
0,1, \ldots, D-1, \quad a \,=\,0,1, \ldots, D-1.
\end{equation}
The variations $\delta ds$ and $\delta k_1(s)$  encountered in
(2.4) can be expressed
now in terms of the functions $\varepsilon ^a (s)$ from (2.11). For
variation $\delta ds$ one obtains
\begin{equation}
\delta \,ds\,=\,\delta\,\sqrt{dx_\mu \,dx^\mu }\,=\,\frac{dx_\mu \,\delta\,
d x^\mu}{ds}\,=\,\dot x_\mu\,d\,(\delta x^\mu )\,{.}
\end{equation}
Substituting (2.12) into (2.4) and integrating by parts, the second
term in (2.4) acquires the form
$$
\delta S_2\,=\,-\int\!\!d\left ({\cal L}(k_1)\,\dot x_\mu \right )
\,\delta x^\mu \,=
$$
\begin{equation}
=\,-\int \!\!\rl '(k_1)\,\dot k_1\,(\dot x_\mu \delta x^\mu )\,ds\,-
 \int \!\!
\rl (k_1)\,(\ddot x_\mu \delta x^\mu )\,ds\,{.}
\end{equation}
By making use of eqs.~(2.8) -- (2.11) we obtain for the variation
$\delta S_2$
\begin{equation}
                   \delta S_2 \,=\,- \int \!\! \rl ' (k_1)\,\dot k_1\,
 \varepsilon ^0(s)\,ds \,- \int \!\! \rl (k_1)\,k_1(s)\,
 \varepsilon ^1(s)\,ds\,{.}
\end{equation}

Now we proceed to calculate  the variation $\delta k_1$. From the definition
 (2.3)
we deduce
\begin{equation}
k_1(s)\,\delta k_1(s)\,=\,-\,\left (  \ddot x _\mu\, \delta\ddot
x^\mu \right )\,=\,
-\,\left (  \dot e_{0 \mu} \,\delta\ddot x^\mu \right )\,{.}
\end{equation}
Here it should be taken into account that the operations  $\delta  $
and $d/ds$ do not commute. The direct calculation shows that
     \begin{equation}
\left [  \delta,\,  \frac  {d}{ds}\right  ]\,=\,-\,  \left ( e_{0\mu
}\, \frac {d}{ds}\delta x^\mu \right )\, \frac{d}{ds}\,{.}
     \end{equation}
Applying this formula one finds
     \begin{equation}
     \delta \ddot x^\mu\,=\, \frac {d^2}{ds^2}\,\delta x^\mu \,
-\,2\,\ddot x^\mu \left (e_{0\nu }\, \frac{d}{ds}\,\delta x^\nu \right )
\,-\,e_0^\mu\,\frac{d}{ds}\,\left ( e_{0\nu }\,\frac{d}{ds}\,
\delta x^\nu \right )\,{.}
     \end{equation}
Substituting (2.17) into (2.15) and taking into account the
Frenet equations (2.9) we obtain
\begin{equation}
\delta k_1(s)\,=\,\varepsilon ^0\dot k_1\,-\,\ddot \varepsilon ^1\,
+\,\varepsilon  ^1 (k^2_1\,+\,k_2^2)\,-\,2\,\dot \varepsilon ^2\,k_2\,
-\,\varepsilon ^2\,\dot k_2\,-\,\varepsilon ^3\,k_2\,k_3\,{.}
\end{equation}
Thus, the variation of the first curvature of the curve,
$k_1(s)$ depends on the
variations of the worldline coordinates only along the directions
$e_0^\mu,\;e_1^\mu,\;e_2^\mu $, and $e_3^\mu $ (on the functions
$\varepsilon ^a (s),\; a \,=\,0,1,2,3 )$ and on the
first three curvatures
$k_1,\;k_2$ and $k_3$.

Substituting (2.18) into (2.4), integrating by parts, and taking
into account (2.14)
we obtain
$$
\delta S \,=\,\int \!\! ds\, \left \{ \left [ \left
(k_1^2\,+\,k_2^2\right ) \,\rl '(k_1)\,
-\,\frac{d^2}{ds^2}(\rl '(k_1))\,-\,k_1\,\rl(k_1)  \right ]\,
\varepsilon ^1(s)\,+
 \right .
$$
\begin{equation}
+\, \left .\left [ 2\,\frac{d}{ds}(\rl '(k_1)\,k_2)\,-\,\dot k_2\,\rl '(k_1)
 \right ] \,\varepsilon ^2(s)\,-\,\rl '(k_1)\,k_2\,k_3\,\varepsilon ^3(s)
 \right \}\,=\,0\,{.}
\end{equation}
 The functions $\varepsilon
^i(s),\;i\,=\,1,2,3$
are arbitrary; therefore we deduce from (2.19) three equations
\begin{eqnarray}
\frac{d^2}{ds^2}\,(\rl'(k_1)) & =& \left ( k_1^2\,+\,k_2^2\right )\,
\rl '(k_1)\,-\,k_1\,\rl(k_1)\,{,}\\
2\,\frac{d}{ds}\,(\rl'(k_1)\,k_2) &=&\dot k_2\,\rl'(k_1)\,{,} \\
\rl '(k_1)\,k_2\,k_3&=&0\,{.}
\end{eqnarray}

A set of equations equivalent to eqs.~(2.20) -- (2.22) has been derived
in the book [23] (see also [26]) by making use of rather complicated
formal mathematical methods, based  on the exterior Cartan forms. We
consider that  the use of simple Frenet equations familiar  to physicists is
enough.

To satisfy eq.~(2.22) one has to put $k_3\,=\,0$. From here it follows
that all the higher curvatures $k_4,\,k_5,\, \ldots, ,\,k_{D-1}$ vanish
also [23]. Thus we have for arbitrary $D$
\begin{equation}
k_n(s)\,=\,0,\quad n\,=\,3,4,\ldots ,D-1\,{.}
\end{equation}

Equation (2.21) can be explicitly integrated. Namely, rewriting this
 equation in the form
$$
2\,\rl ''(k_1)\,k_2\,dk_1\,+\,\rl '(k_1)\,dk_2\,=\,0
$$
we obtain
\begin{equation}
2\, d\, \left [\ln \rl '(k_1)\right ]\,+\,d\,(\ln k_2)\,=\,0\,{.}
\end{equation}
Therefore,
\begin{equation}
\left ( \rl ' (k_1)\right )^2k_2\,=\,C\,{,}
\end{equation}
where $C$ is an integration constant.

 Eq.~(2.25) allows us to   eliminate   $k_2(s)$ from
eq.~(2.20). As a result, one non-linear equation of the second order for
the worldline curvature, $k_1(s)$, arises
\begin{equation}
\frac{d^2}{ds^2}\,(\rl '(k_1))\,=\,\left ( k_1^2 \,+\,
\frac{C^2}{(\rl '(k_1))^4}\right ) \,\rl '(k_1) \,-\,k_1\,\rl (k_1)\,{.}
\end{equation}

Thus, we have derived a complete set of equations (see eqs.~(2.23),
(2.25), and (2.26)) for principal curvatures $k_i(s),\;i\,=\,1,2, \ldots
, D-1 $ of the worldline of the particle.

\section{Integrability of the Euler-Lagrange equations
for principal curvatures}
\setcounter{equation}0

It is remarkable that a first integral for  eq.~(2.26)  can be find in
a general form for arbitrary Lagrangian $\rl (k_1)$. This first
integral naturally arises  analysing  the
Euler-Lagrange equations written in
terms of the worldline coordinates  $x^\mu (s)$. The
constants of integration turn out  to be the particle mass and its spin.

In this section we shall use an arbitrary parametrization of
the worldline $x^\mu (\tau )$.  From now on, a dot over $x$ will denote the
differentiation with respect to evolution parameter $\tau$; prime
on the Lagrangian function $\rl $ will, as before,
denote the differentiation with respect to its
argument $k_1$. The
action (2.1) assumes  the form
\begin{equation}
S\,=\,\int \!\! L(\dot x,\,\ddot x)\,d \tau \,=\, \int \!\!\rl (k_1)\,
\sqrt {\dot x^2}\,d \tau\,{,}
\end{equation}
where
\begin{equation}
L(\dot x,\,\ddot x)\,=\,\sqrt {\dot x ^2}\,\rl (k_1)\,{,}
\end{equation}
\begin{equation}
k_1^2\,=\,\frac{(\dot x\ddot x)^2\,-\,\dot x^2\,
\ddot x^2}{(\dot x^2)^3}\,{.}
\end{equation}

Introducing the conserved Lorentz vector of  energy-momentum
\begin{equation}
P^\mu \,=\,\frac{d}{d \tau}\,\left (  \frac{\partial L}{\partial
\ddot x_\mu }
\right ) \,-\, \frac{\partial L}{\partial \dot x _\mu }\,{,} \quad
\mu \,=\,0,1, \ldots, D-1,
\end{equation}
the Euler-Lagrange equations generated  by (3.1) are written as
\begin{equation}
\frac{d}{d \tau}\,P^\mu \,=\,0, \quad \mu \,=\,0,1,\ldots , D-1.
\end{equation}
Now we pass in eq.~(3.4) from an arbitrary evolution
variable $\tau $ to the
natural parameter $s$, using the following relations
\begin{equation}
\frac{d}{d\tau }\,=\,\sqrt {\dot x^2}\,\frac{d}{ds\,}{;}\qquad
\frac{d^2 x_\mu}{ds^2\,}
 \,=\,\frac{\dot x^2\ddot x_\mu \,-\,(\dot x \ddot x)\,
\dot x_\mu}{(\dot x^2)^2}\,{;}
\qquad k_1\,\frac{\partial k_1}{\partial \ddot x_\mu}\,=\,
-\,\frac{1}{\dot x^2}\,\frac{d^2 x_\mu}{ds^2\,}{;}
\end{equation}
$$
k_1\,\frac{\partial k_1}{\partial \dot x_\mu }\,=\, \frac{1}{(\dot x^2)^4}\,
\left \{ \dot x^2\,(\dot x \ddot x)\,\ddot x^\mu \,+\,\left [
2\, \dot x^2\, \ddot x^2\,-\,3\,(\dot x \ddot x)^2\right ]
\dot x^\mu \right \}
$$
As a  result, the energy-momentum vector acquires the form
\begin{equation}
P^\mu \,=\,\frac{dx_\mu}{ds\,} \,(2\,\rl '\,k_1\,-\,\rl )\,+
\,\frac{d^2 x_\mu}{ds^2\,}
\frac{dk_1}{ds\,}\,\left (
 \frac{\rl '\,}{k_1^2}\,-\,\frac{\rl ''}{k_1}\right )\,
-\,\frac{\rl '}{k_1}\,\frac{d^3 x_\mu}{ds^3\,}\,{.}
\end{equation}

Further we shall use the direct consequences of the definitions
 (2.2) and (2.3)
\begin{equation}
\frac{dx_\mu}{ds\,}\frac{d^2 x_\mu}{ds^2\,}\,=\,0; \qquad
\frac{dx_\mu}{ds\,}\frac{d^3 x_\mu}{ds^3\,}\,=\,k_1^2;\qquad
\frac{d^2 x_\mu}{ds^2\,}\frac{d^3 x_\mu}{ds^3\,}\,=
\,-\,k_1\,\frac{dk_1}{ds\,}\,{.}
\end{equation}
The second curvature (or torsion), $k_2$,  is defined in the following
way [27]
\begin{equation}
k_1^4\,k_2^2\,=\,{\det}_G\left (\frac{dx_\mu}{ds\,},\frac{d^2 x_\mu}{ds^2\,},
\frac{d^3 x_\mu}{ds^3\,}\right ),
\end{equation}
where ${\det}_G\left ({dx_\mu}/{ds\,},{d^2 x_\mu}/{ds^2\,},
{d^3 x_\mu}/{ds^3\,}\right )$ is the Gramm determinant for vectors
${dx_\mu}/{ds\,},{d^2 x_\mu}/{ds^2\,},
{d^3 x_\mu}/{ds^3\,} $ [28]. It enables one to express $
 ({d^3 x_\mu}/{ds^3\,})^2$ in terms of $k_1$ and $k_2$
\begin{equation}
\left (
\frac{d^3 x_\mu}{ds^3\,}\right )^2\,=\,k_1^4\,-\,\left(\frac{dk_1}{ds\,}
\right)^2\,-\,k_1^2\,k_2^2\,{.}
\end{equation}
Bearing in mind eq.~(2.25), the torsion $k_2$ can be eliminated from (3.10)
\begin{equation}
\left(\frac{d^3 x_\mu}{ds^3\,}\right)^2\,=\,k_1^4\,-
\,\left(\frac{dk_1}{ds\,}\right)^2\,
-\,k_1^2\,\frac{C^2}{(\rl' (k_1))^4}\,{.}
\end{equation}

Now we square the right  and the left hand sides of eq.~(3.7) and take
into account (3.8) and (3.11). As a result, we obtain
\begin{equation}
M^2\,\equiv \,P^2\,=\,\rl^2\,-\,\left (\frac {d}{ds} \rl '\right )^2
\,-\,2\,\rl\,\rl 'k_1\,+\,(\rl ')^2\,k_1^2\,-\,
\frac{C^2}{(\rl ')^2}\,{.}
\end{equation}
It turns out that eq.~(3.12) is the first integral for
the Euler-Lagrange equation (2.26)
derived in the previous Section.
If $\rl ''\, \not= \, 0$, one can  be
convinced of this relevant result by  direct differentiation of (3.12)
 with respect to $s$.
For Lagrangians linear in $k_1$ equations (2.26) and (3.12) should be
 treated as independent ones
because the differentiation of (3.12) identically gives zero (see
Appendix A).

It follows that the order of eq.~(2.26) can be reduced by one.
 From (3.12) we obtain
\begin{equation}
\frac{dk_1}{ds}\,=\,\pm\,\sqrt{f(k_1)}\,{,}
\end{equation}
where
\begin{equation}
f(k_1)\,=\,\frac{1}{(\rl'')^2}\,\left \{\rl ^2\,-\,2\,\rl \,\rl ' \,k_1\,
+\,(\rl ')^2\,k_1^2 \,-\,\frac{C^2}{(\rl ')^2}\,-\,M^2\right \}\,{.}
\end{equation}
Integration of (3.13) gives
\begin{equation}
\int\limits_{k_{10}}^{k_1}\!\!\frac{dx}{\sqrt{f(x)}}\,
=\,\pm\,(s\,-\,s_0)\,{,}
\end{equation}
where $k_{10}\,=\,k_1(s_0)$.

Thus, formula (3.15) defines the curvature of the worldline as a function
of $s$. Equations (2.23) and (2.25) determine
the remaining curvatures. As a result, for arbitrary Lagrangian
$\rl (k_1)$, the problem of finding the principal curvatures
of the worldline is reduced to quadratures.

Another interesting result can be obtained  from the study of
eqs.~(3.4), (3.5), that enables us to derive the
relation (2.25) between curvature and torsion  in a new way.
 It turns out that the integration constant $C$ can be expressed
 in terms of the
particle spin and mass.\footnote{In the models defined
by action (2.1) spin of the particle proves to be non zero
at the classical level already.
Its  value is ultimately determined by the initial
conditions for corresponding equations of motion.}
Let us show this explicitly.

The invariance of the action (2.1) under Lorentz
transformations entails the conservation of the angular-momentum tensor
\begin{equation}
M_{\mu \nu}\,=\,\sum_{a=1}^{2}(q_{a\mu}p_{a\nu}\,-\,q_{a\nu}p_{a\mu})\,{,}
\end{equation}
where the canonical variables  $q_a^\mu$ and $p^\mu _a$ are
defined as follows
$$
q_{1\mu }\,=\,x_\mu, \quad q_{2\mu}\,=\,\dot x_\mu,
$$
\begin{equation}
p_{1\mu}\,=\,P_\mu\,=\,-\,\frac{\partial L}{\partial
\dot x_\mu}\,-\,\frac{dp_2}{d \tau}, \quad p_{2\mu }
\,=\,-\,\frac{\partial L}{\partial \ddot x_\mu}\,{.}
\end{equation}
The spin $S$ of the particle will be also a
conserved quantity. In the case of $D$-dimensional
space-time spin $S$ is defined by [29]
\begin{equation}
S^2\,=\,\frac{W}{M^2}\,{,}
\end{equation}
where
$$
W\,=\,\frac{1}{2}\,M_{\mu \nu}M^{\mu \nu}p_1^2\,-
\,(M_{\mu \sigma}p_1^\mu )^2,\quad
M^2\,=\,p_1^2\,{.}
$$
For $D\,=\,4$ the invariant $W$ is the squared
Pauli-Lubanski vector with sign minus
\begin{equation}
W\,=\,-\,w_\mu w^\mu ,\quad w_\mu \,=\,\frac{1}{2}\,
\varepsilon _{\mu \nu \rho \sigma} M^{\nu \rho} p_1^\sigma \,{.}
\end{equation}
 From these equations it follows
\begin{equation}
M^2\,S^2\,=\,k_2^2(\rl ')^4\,{.}
\end{equation}
Hence, the integration constant $C$ in (2.25) is given by
\begin{equation}
C^2\,=\,M^2\,S^2\,{.}
\end{equation}
After quantization, $S^2$ in (3.21) should be replaced
in the following way
\begin{equation}
S^2\,\to \,S\,(S\,+\,D\,-\,1), \quad S\,=\,0,1, \ldots ,
\end{equation}
where $D$ is the dimension of the space-time.

Therefore, dealing with the Euler-Lagrange equations written
 in terms of the  curve coordinates $x^\mu(s)$, we  have derived  the basic
equations (3.12) and (3.20) for the principal curvatures $k_1$
and $k_2$ and have specified the integration constants.\footnote{In paper
[4] the Euler-Lagrange equations in terms of $x^\mu (s)$ have been used
for obtaining some auxiliary  conditions that should
be satisfied by $k_1(s)$.}
 Probably,  the  equations
(2.23) can be derived in this way also. However, checking this possibility
is beyond the scope of our consideration.

In conclusion of this section a general note concerning the mass
spectrum in the models in question
should be done. From eqs.~(3.12) and (3.21) it follows that
the  tachyonic states with $M^2\,<\,0$ are unavoidably  present
in these theories already at the
classical level [30]. It takes place at any values of the particle spin $S$.
\section{Particle with maximal proper acceleration}
\setcounter{equation}0

We apply the approach developed above to the
model of a relativistic particle with
maximal proper acceleration. The Lagrangian of this
model is uniquely determined by a modified form of the infinitesimal
space-time interval
\begin{equation}
d\tilde \sigma^2 \,=\,\left (1\,-\,L_0^2\,
\left ( \frac{d^2x_\mu}{ds^2}\right )^2
\right )\, ds^2\,{,}
\end{equation}
where $L_0$ is a fundamental constant with dimension of length. The physical
 motivation for considering such a metric form can be found in papers
 [13 -- 18]. Here
we make  only  a short comment: mathematically, the introduction of the
line element (4.1) (the metric structure of the second order)
means that
submanifolds in the Minkowski space-time are now  Kawaguchi spaces
[31] rather than  Riemannian spaces.  In particular, the worldline
should be geometrically treated as a one-dimensional Kawaguchi space.
 In pure geometry the Kawaguchi spaces
are considered for a long time [32, 33].

Substituting the interval $ds$ in a standard action for spinless
relativistic particle
\begin{equation}
S_0\,=\,-\,m\int \!\! ds
\end{equation}
by $d\tilde \sigma$ we obtain
\begin{equation}
S\,=\,-\,\mu _0\int \!\!\sqrt{M_0^2\,-\,k_1^2}\,ds,
\end{equation}
where $\mu _0\,=\,m/M_0,\;M_0\,=\,L_0^{-1}$. When $M_0^2\,\to\,
\infty $ the action (4.3) reduces to (4.2).

Obviously, the investigation of the Euler-Lagrange equations for
action (4.3) in terms of particle coordinates
 is practically hopeless task. However, by making use of
the  method developed above, we can easily show that the
acceleration is actually superioly limited by $M_0^2$.
We shall make use  of  eqs.~(3.13)
-- (3.15), substituting there
\begin{equation}
\rl (k_1)\,=\,-\,\mu _0 \,\sqrt{M_0^2\,-\,k_1^2}\,{.}
\end{equation}
Integral in eq.~(3.15) can be easily expressed in terms of the elementary
functions. However, the final formula turn out to be
rather complicated. For simplicity, we exactly integrate eq.~(3.13) for
 particle spin $S$ equal to zero.
 From eqs.~(3.13), (3.14) and (4.4) we deduce
\begin{equation}
\left ( \frac{dk}{d\bar s}\right )^2\,=\,(1\,-\,k^2)\,[1\,
-\,\mu ^{-2}\,(1\,-\,k^2)]\,{,}
\end{equation}
where $k(\bar s)$  is the dimensionless proper acceleration,
$k^2\,=\,k_1^2/M^2_0$, and
$\bar s\,=\,M_0\,s$ is the dimensionless length of   worldline.
Here a new parameter $\mu$
is introduced, $\mu ^2 \,=\,(m/M)^2$, where $M$ is the particle mass
$M^2\,=\,P^2$. We confine ourselves to
a positive $M^2$ (tachyonic solutions are not considered).

Obviously, the initial values for $k(\bar s)$
should belong to the interval
\begin{equation}
0\,\leq \,k^2 \,< \,1 .
\end{equation}
Therefore in the solutions of the equation (4.5)
we have to take only branches
that have    common points  with the interval (4.6). Depending on the
value of the parameter $\mu ^2$ we have here two possibilities.

If
\begin{equation}
\mu ^2\,=\,\frac{m^2}{M^2}\,\leq\,1\,{,}
\end{equation}
then $k^2(\bar s)$ varies in the region
\begin{equation}
1\,-\,\mu ^2\,\leq\, k^2\,<\,1
\end{equation}
and it is defined by
\begin{equation}
k^2(\bar s)\,=\,1\,-\,\mu ^2\,\frac{(\coth \bar s)^2\,
-\,1}{(\coth \bar s)^2\,-\,\mu^2}\,{.}
\end{equation}
The constant $s_0$ in eq.~(3.15) is chosen in such a way that
\begin{equation}
k^2(0)\,=\,1\,-\,\mu ^2\,{.}
\end{equation}
 From (4.9) it follows that
\begin{equation}
k^2(\bar s)\,\to\,1^-\,
\end{equation}
when $\bar s \,\to \, \pm\, \infty $.
 If
\begin{equation}
\mu ^2\,=\,\frac{m^2}{M^2}\,>\,1\,{,}
\end{equation}
then $k^2$ varies in the region
\begin{equation}
0\,\leq\,k^2\,<\,1
\end{equation}
and it is defined by
\begin{equation}
k^2(\bar s)\,=\,1\,-\,\mu ^2\,\frac{(\tanh \bar s)^2\,
-\,1}{(\tanh \bar s)^2\,-\,\mu^2}\,{.}
\end{equation}
The boundary values of $k$ in this case are
\begin{equation}
k^2(0)\,=\,0, \quad k^2(\bar s)\,\to \,1^-\,
\mbox{ if }\bar s \,\to \,\pm\,\infty\,{.}
\end{equation}

The behaviour of $k^2(\bar s)$ near 1 can be easily investigated in the
general  case of nonzero spin $S$.  To this end an approximate
differential  equation  for  $k^2(\bar  s)$  should  be   used.   If
$k^2\,\to\,1^-\,$ then we get from eqs.~(3.13), (3.14) and (4.4)
\begin{equation}
\left ( \frac{dk}{d\bar s} \right )^2\, \approx \,1\,-
\,k^2, \quad k^2\,\to \,1^-\,\,{.}
\end{equation}
Integration of this equation in the range of $k^2$ under consideration gives
\begin{equation}
k^2(\bar s) \,\approx \, \tanh ^2(\bar s \,-\,\bar s_0), \quad
\bar s\, \to \,\pm \, \infty\,{.}
\end{equation}
Hence, the proper acceleration of
the particle obeys the restriction
\begin{equation}
k^2(\bar s)\,< \,1\,{.}
\end{equation}

In conclusion of this section it should be noted
the following. In the models with the Lagrangians (see papers [4, 30, 34
-- 36])
\begin{eqnarray}
\rl (k_1)&=&-\,m\,-\,\alpha\,k_1(s)\,{,}\\
\rl (k_1)&=&-\,\alpha \,k_1(s)
\end{eqnarray}
the particle  spin  $S$  and its mass $M$ are not independent;  they
have to obey a mass-spin relation determining the Regge  trajectory
(see Appendix A).
In  the  model under consideration with action (4.3) $M$ and $S$ are
arbitrary independent  constants  of  motion.  This  distinction  is
easily explained  by  different  numbers  of constraints in the phase
space in these models.  The point is that all of  these  Lagrangians
are  singular  or degenerate ones and as a result the  corresponding
  phase spaces are
restricted by  constraints.  In  the  model  (4.19)  there  are  four
Hamiltonian  constraints  and in the model (4.20) the number of such
constraints  is  five.  In  the  model  (4.3)  there  are  only  two
constraints  (see  Appendix  B where the Hamiltonian formalism for this
model is constructed).  This means that the models (4.19) and (4.20)
possess   extra  symmetries  in  addition  to  the  reparametrization
invariance and such symmetries are absent in the model (4.3).

\section{Conclusion}
     The formulae  (2.20)  -- (2.26),  (3.13) -- (3.15),  and (3.17)
provide a  complete  solution  to  the  problem  of  obtaining
equations  of  motion for arbitrary Lagrangians $\rl (k_1)$ in
terms of the principal curvatures of  the  worldline,  and  of
integrating  these  equations  by quadratures;  moreover,  the
integration constants are expressed in terms of  the  particle
mass $M$ and its spin $S$.  For comparison sake,  we note that
in the book [23] and in paper [26] the complete  integrability
of  the  equations  of  motion  was  been  proved only for the
simplest Lagrangians  $\rl  (k_1)$  linear  and  quadratic  in
$k_1$,   without  elucidating  the  physical  meaning  of  the
integration constants.

In present paper we do not touch upon the
problem of recovering the world trajectory  by  means  of  its
principal curvatures. If all the curvatures $k_i,\;i\,=\,
1,2,  \ldots,  D-1$ are constants, then this problem can be solved easily
(see,  for  example,  [27]). Obviously, in general
case one encounters difficulties.
  However,  it  seems to us that the key
properties of the dynamics in the  models  under  consideration  are
determined   fully  enough.  We  have  relate  the  physical
characteristics of the particle (its mass and spin) with geometrical
invariants (principal curvatures) of its world trajectory.

     As regards  the model (4.3), it  gives us nontrivial  example  of
consistent relativistic dynamics obeying the principle of superiorly limited
 proper acceleration.
\vskip1.4cm
{\Large \bf Acknowledgement}
\vskip1.0cm

The basic part of this paper has been accomplished during the stay of
one  of the authors, V.\ V.\ N., at the Salerno University. Taking an
opportunity he would like to thank Prof.\ G.\ Scarpetta and his
collaborators for their kind hospitality. Some topics concerned in paper
have been discussed with Dr.~H.~Arod\'z (Jagellonian University,
Cracow). V.~V.~N.~is grateful to him for this.

\appendixa

Here Lagrangians  linear in $k_1$ (see eqs.\ (4.19) and (4.20))
will  be  considered  in  the  framework  of  the  geometrical
approach developed in the present paper. For Lagrangian (4.19)
equations (2.20) and (3.21) give
\begin{equation}
             k_1^2\,(1\,-\,\alpha )\,+\,k_1\,m\,+\alpha
             ^{-3}M^2S^2\,=\,0\,{.}
\end{equation}
Hence, curvature of the world trajectory should be a constant expressed
in terms of the particle mass $M$ and spin $S$ and the parameters of the
model $m$ and $\alpha $. In view of this equation (3.12) reduces to the
following mass-spin relation
\begin{equation}
M^2\,=\,\frac{m^2}{1\,+\,\alpha ^{-2}S^2}{.}
\end{equation}
Transition to the quantum theory can be made by the substitution (3.22).
Therefore, we obtained the basic results concerning this model without
dealing with the constraints that appear in the Lagrangian or Hamiltonian
treatments of this model.

In the case of the Lagrangian  (4.20)  we easily deduce from eq.~(2.21)
that the torsion $k_2$ in this model should be a constant. Then
it follows from (3.12) and (3.20) that the mass and spin   are
equal to zero. Equation (3.12) can be rewritten as
\begin{equation}
P^2\,=\,-\,\alpha ^{-2}C^2\,=\,\alpha ^{-2}\,W\,{.}
\end{equation}
It means  that Pauli-Lubanski vector $w^\mu $ is now isotropic
as well as $P^\mu$  and  they  are  proportional  each  other.
Therefore,  parameter  $\alpha $ is a helicity of the massless
particle.  In  quantum  case  it  takes   only   integer   and
half-odd-integer values.

\appendixb
Here we shortly consider  the Hamiltonian formalism for the model (4.3).
Canonical variables $q^\mu_a,\;p^\mu_a,\;a\,=\,1,2$ are introduced
according to eqs.~(3.17). By making use of the definition of $p_2^\mu$ in
(3.17) and the explicit form for $L$ in (4.3) we find primary constraint
\begin{equation}
\varphi(q,\,p)\,=\,p_{2\mu }\,q_2^\mu\,\approx \,0\,{,}
\end{equation}
where $\approx  $ means a weak equality [33]. There are no other primary
constraints in the model under consideration. It is an essential
distinction of this model as compared with the models (4.19) and (4.20).

Canonical Hamiltonian is introduced in the standard way
\begin{equation}
H\,=\,-\,p_1^\mu \,\dot x_\mu \,-\,p_2^\mu \,\ddot x_\mu \,-\,L\,{.}
\end{equation}
In terms of the canonical variables $H$ assumes the form
\begin{equation}
H\,=\,-\,p_1q_2\,+\,M_0\,\sqrt{q_2^2\,(\mu ^2_0\,-\,q_2^2p_2^2)}\,{.}
\end{equation}
The dynamics in the phase space is determined by the total Hamiltonian
\begin{equation}
H_{\rm T}\,=\,H\,+\,\lambda(\tau )\,\varphi (q,\,p)\,{,}
\end{equation}
where $\lambda (\tau )$ is the Lagrange multiplier. The equations of
motion in the phase space are written as follows
\begin{equation}
\frac{df}{d \tau}\,=\,\frac{\partial f}{\partial \tau}\,+\,\left \{
f,\, H_{\rm T} \right \}\,{,}
\end{equation}
where $f$ is an arbitrary function of the canonical
variables and evolution parameter $\tau $ and $\{\ldots ,
\, \ldots \}$ stands for the Poisson brackets
\begin{equation}
\{f,\,g\}\,=\,\sum_{a=1}^{2}\left ( \frac{\partial f}{\partial p_a^\mu }
\,\frac{\partial g }{\partial q_{a \mu}}\,-\,
 \frac{\partial f}{\partial q_a^\mu }
\,\frac{\partial g}{\partial p_{a \mu}} \right )\,{.}
\end{equation}

Requirement of the stationarity of the primary constraint (A.1)
\begin{equation}
\frac{d \varphi }{d\tau }\,=\,\left \{ \varphi, \,H_{\rm
T}\right \}\,\approx \,0
\end{equation}
results in secondary constraint
\begin{equation}
\left \{ \varphi, \,H_{\rm T}\right \}\,=\,
\left \{ \varphi, \,H\right \}\,=\, H\,\approx \,0\,{.}
\end{equation}
As one would expect, the canonical Hamiltonian in the model in question
vanishes in a weak sense. It is  actually a
consequence of the reparametrization
invariance of the initial action (4.3).

Finally, in the model (4.3) there are only two constraints in
the phase space, $\varphi $ and
$H$. The Hamiltonian formalism presented here can be used  as the
basis of the canonical quantization of the model under consideration.

     \end{document}